\documentclass[usenatbib]{mnras}

\usepackage[T1]{fontenc}
\usepackage{ae,aecompl}

\usepackage{verbatim} 
\usepackage{xcolor}
\usepackage{subfig}
\usepackage{lipsum}

\usepackage{graphicx}	
\usepackage{amsmath}	
\usepackage{amssymb}	

\title[Mass Segregation-Concentration Correlation in Globular Clusters]{Correlation Between Mass Segregation and Structural Concentration in Relaxed Stellar Clusters}

\author[de Vita et al.]{
Ruggero de Vita$^{1}$\thanks{E-mail: r.devita@student.unimelb.edu.au},
Michele Trenti$^{1}$
and Morgan MacLeod$^{2}$\thanks{NASA Einstein Fellow}.
\\
$^{1}$School of Physics, The University of Melbourne, VIC 3010, Australia\\
$^{2}$Harvard-Smithsonian Center for Astrophysics, 60 Garden Street, Cambridge, MA 02138, USA\\
}

\date{Accepted 2019 March 17. Received 2019 March 9; in original form 2019 January 16}

\pubyear{2018}

\begin{document}
\label{firstpage}
\pagerange{\pageref{firstpage}--\pageref{lastpage}}
\maketitle

\begin{abstract}
The level of mass segregation in the core of globular clusters has been previously proposed as a potential indicator of the dynamical constituents of the system, such as presence of a significant population of stellar-mass black holes (BHs), or even a central intermediate-mass black hole (IMBH). However, its measurement is limited to clusters with high-quality Hubble Space Telescope data. 
Thanks to a set of state-of-the-art direct N-body simulations with up to 200k particles inclusive of stellar evolution, primordial binaries, and varying BH/neutron stars, we highlight for the first time the existence of a clear and tight linear relation between the degree of mass segregation and the cluster structural concentration index. The latter is defined as the ratio of the radii containing 5\% and 50\% of the integrated light ($R_5/R_\mathrm{50}$), making it robustly measurable without the need to individually resolve low-mass stars. Our simulations indicate that given $R_5/R_\mathrm{50}$, the mass segregation $\Delta m$ (defined as the difference in main sequence median mass between center and half-light radius) is expressed as $\Delta m/M_\odot = -1.166 R_5/R_\mathrm{h} + 0.3246$, with a root-mean-square error of $0.0148$. In addition, we can explain its physical origin and the values of the fitted parameters through basic analytical modeling. 
Such correlation is remarkably robust against a variety of initial conditions (including presence of primordial binaries and IMBHs) and cluster ages, with a slight dependence in best-fit parameters on the prescriptions used to measure the quantities involved. Therefore, this study highlights the potential to develop a new observational tool to gain insight on the dynamical status of globular clusters and on its dark remnants. 
\end{abstract}



\section{Introduction} \label{sec:intro}








Old ages and high stellar densities make globular clusters (GCs) natural laboratories for a range of diverse astrophysical processes \citep{heggie-hut}. In fact, their current stellar populations are the manifestation of more than 10 Gyr of combined stellar, dynamical, and hydrodynamical evolution, whose interplay is primarily responsible for enhanced presence of exotic objects \citep[e.g.][]{bailyn:95}, including blue stragglers stars (see e.g. \citealt{ferraro:97,lanzoni:07}) and binary pulsars \citep{camilo:00,benacquista:13}. GCs have also been indicated as possible formation sites of intermediate-mass black holes (IMBHs, \citealt{portegies:02,vesperini:10}) and as favourable environments for mergers of dark compact objects \citep[e.g.][]{2014ApJ...784...71S,abbott:16a,macleod:16,rodriguez:16,2017ApJ...846...36S,2018ApJ...853..140S}. 

However, understanding the dynamical evolution of GCs remains challenging, in particular with respect to the presence and contribution of dark constituents (BHs, neutron stars) which cannot be observed directly (see e.g. \citealt{noyola:08,lanzoni:13,lutzgendorf:15}). In this context, a variety of different tracers for the BHs presence have been proposed, with one of them relying on the characterization of the long-term dynamical evolution of GCs \citep{gill:08,pasquato:09,trenti:13,peuten:16,bianchini:17,weatheford:17,arca-sedda:18,askar:18}, which is driven by the tendency of the system to evolve towards a state of energy equipartition through two-body relaxation (see \citealt{binney}). 

Because GC constituents have a spectrum of masses, the evolution toward (partial) energy equipartition causes the system to become spatially mass segregated. More massive objects become preferentially restricted toward the minimum of the gravitational potential (the cluster's central region) as they trend toward energy equipartition with lighter counterparts through gravitational two-body encounters. Correspondingly, the velocity dispersion and spatial extent of lighter-than-average stars increases. Complete energy equipartion is never reached (see \citealt{trenti:13,bianchini:16,spera:16}) because GCs are open systems -- stars that gain too much energy become unbound. 
The two-body relaxation time (which has a median value of $\approx 10^8$ yr for galactic GCs; \citealt{heggie-hut}) is the relevant time-scale over which globular clusters undergo mass segregation, and it depends primarily on mass and radius, with compact low-mass clusters characterized by shorter relaxation times compared to extended and high-mass counterparts.

In this context, it is well established by early numerical modeling studies that massive dark remnants, and in particular IMBHs, affect mass segregation and energy equipartition, specifically quenching them  \citep{trenti:07,gill:08,pasquato:09,trenti:13}. Partial suppression of mass segregation has also be shown to be induced by a population of primordial binaries \citep{gill:08,beccari:10,pasquato:16,webb:17}, as well as by a significant presence of stellar BHs \citep{alessandrini:16,peuten:16,baumgardt-sollima:17,weatheford:17}. 
These three different classes of objects are thought to act (either in combination or individually) by enhancing strong three-body scattering events in the GC core, which enhance the probability of imparting significant kicks to objects interacting with them almost independently of their mass. This partially redistributes core objects (preferentially more massive on average) throughout the system, thus reducing the amount of mass segregation and energy equipartion \citep{trenti:13}.

These theoretical/numerical studies suggest that the level of mass segregation (and energy equipartion) in a GC can thus be used to infer useful information on its dynamical state and (dark) constituents. However, the measurement of such quantities is observationally challenging, mainly because it requires sufficiently high resolution images of the central crowded regions of star clusters that can resolve individual stars of low mass. While feasible and demonstrated for GCs such as NGC2298 \citep{pasquato:09}, M10 \citep{beccari:10}, as well as for Omega Centauri \citep{anderson:10,trenti:13}, in practice observational limitations restrict the mass segregation/energy equipartition dynamical analysis to the subset of galactic GCs that have relatively low densities and high quality Hubble-Space-Telescope photometry in multiple bands/epochs (see e.g., \citealt{beccari:10,bellini:14,webb:17b,libralato:18}). 

Such observational challenges highlight the need for alternative observables that can be used to characterize the dynamical state of GCs (see e.g. \citealt{bianchini:16}) and, in turn, to infer the properties of the dynamical constituents of a broader range of galactic and extragalactic GCs. To this purpose,
we analyze in this study a large set of realistic direct N-body simulations of star clusters, which includes a variety of different initial conditions and setups, searching for quantities that correlate with the degree of mass segregation once the system becomes dynamically old (i.e. old compared to its two-body relaxation timescale). We present evidence for a tight correlation between the level of mass segregation of dense stellar systems and their structural concentration, measured in a novel but easy to asses way by considering the ratio of the radii containing 5\% and 50\% of the projected light. The correlation is then tested for robustness against a variety of operational choices for defining mass segregation and concentration, overall demonstrating remarkable resilience and low residuals across the whole set of simulations, despite their significant diversity in initial conditions. 

The paper is organized as follows. In Sec.\ref{sec:simulations} we describe the simulations used in this work and define mass segregation and structural concentration, focusing in particular on prescriptions that can be implemented from actual GC observations. In Sec.\ref{sec:results} we demonstrate the existence of the mass-segregation structural concentration correlation, and test for robustness against different observational and simulation setups. In Sec.\ref{sec:physic_interp} we present a physical interpretation for this correlation by means of a simplified order-of-magnitude model. Finally, we conclude in Sec.\ref{sec:conclusions} with an outlook for future observational testing and applications of this newly discovered tool to investigate GC dynamics.

\section{Methods} \label{sec:simulations}

\subsection{Numerical framework} \label{subsec:setup}
The set of star cluster simulations used in this paper is obtained using the direct N-body integrator NBODY6 \citep{aarseth:03} inclusive of GPU support \citep{nitadori:12}, as well as stellar evolution look-up tables for single and binary stars, through the SSE and BSE packages (originally presented in \citealt{hurley:00}).

The simulated clusters (see Table \ref{tab:simul}) have initial conditions sampled from a \cite{king:66} model distribution with central dimensionless potential $W_0=7$ and half-mass radius $r_\mathrm{h,0} = 2.5$ pc, and up to $N=200000$ particles. 
For the set of simulations considered in this study the adopted initial star distribution represents a fixed condition, which may potentially affect other physical processes such as the kick velocity of dark remnants (see e.g. \citealt{baumgardt:03, contenta:15}) and, in turn, the degree of mass segregation. Despite this numerical limitation related to the computationally challenging task of running large sets of direct N-body models, we expect different values of $W_0$ or $r_\mathrm{h,0}$ to affect only the first stages of the cluster evolution, with differences in the main structural parameters becoming unimportant after a few relaxation times \citep{trenti:10}.

Initial stellar masses are drawn from a \cite{kroupa:01} initial mass function (IMF) in the mass range $0.1\text{-}100\ M_\odot$ irrespective of their radial position in the system. In fact, we do not include primordial mass segregation as its effects on the overall mass function evolution should be lost at later times, when clusters begin losing stars via tidal stripping (see Subsec. 3.5 of \citealt{webb:16}). The simulations include tidal forces, computed assuming that the clusters follow circular orbits in a point-mass galactic gravitational field (see \citealt{trenti-heggie-hut:07} for details), underfilling the tidal radius by a factor of 3. 

A subset of the initial conditions include a central IMBH of $100\text{-}400M_\odot$ which represents $0.15\text{-}0.3$ per cent of the initial cluster mass (see \citealt{macleod:16} and \citealt{devita:18} for details of the setup). The IMBH is initialized with zero velocity at the center of mass of the system, but it is free to wander through the core as a result of dynamical interactions with other constituents. In addition, because of tidal disruption events which follow close encounters with the IMBH, its mass increases during the simulation (generally by $20-40\%$).   

Compared to earlier works that characterized mass segregation (see e.g., \citealt{gill:08}, \citealt{pasquato:09}, \citealt{trenti:13}), we resort to a larger set of realistic simulations that not only include stellar evolution but also have higher number of particles. Since one debated aspect of stellar evolution is the typical velocity distribution of natal kicks imparted to dark remnants (white dwarfs, neutron stars and black holes), we employ different scenarios to investigate systems that include different retention fractions. Specifically, we assign black holes and neutron stars natal kicks drawn from the same Maxwellian distribution with a dispersion $\sigma_*$ of either 1 or 2 times the initial cluster velocity dispersion, $\sigma_*=\sqrt{GM_\mathrm{tot}/r_\mathrm{hm}}$, with $r_\mathrm{hm}$ half-mass radius. No natal kick is given to white dwarfs. While these assumptions do not rely on a specific model of stellar evolution (see \citealt{mirabel:17,mapelli:18} for recent reviews on the topic), the aim of our work is to test GCs' simulations against different retention fractions of stellar remnants, which radically affect the long-term dynamical evolution (see, e.g., \citealt{contenta:15}), and the simple recipe employed is thus sufficient for our scope.

In addition to the retention fraction, the mass spectrum of stellar-mass BHs is also a critical factor influencing the dynamical evolution of the system and the development of mass segregation. Recent development in theories of supernova explosion might suggest that BHs form with masses larger than previously thought (see e.g. \citealt{fryer:12,spera:15}).
To explore different scenarios in regards to this, we generate the initial conditions assuming a different metallicity $Z$, which in turn affects the stellar evolution packages, leading to more massive stellar BHs formed in metal-poor environment (see \citealt{hurley:00}). This way we can effectively simulate conditions where BHs have masses in excess of 20$M_{\odot}$ with the standard (and extensively validated) stellar evolution packages of NBODY6.

Furthermore, our simulations include realizations starting with $1\text{-}10\%$ primordial hard binaries. The initial binary fraction is defined as $ f = 2n_b / (n_s + 2n_b)$, with $n_s$ and $n_b$ number of singles and binaries, respectively. The semi-major axis for each binary pair is computed from a flat distribution in logarithmic space within the range $0.1\text{-}10$ AU, while eccentricities are drawn from a thermal distribution. This particular choice guarantees that most of the binaries ($\gtrsim 80$\%) are not disrupted in the initial stages of evolution (see e.g. \citealt{heggie:75,heggie:06,trenti-heggie-hut:07}).

Finally, the chaotic nature of the N-body problem requires to test the robustness of our results at a statistical level. Therefore, we performed multiple simulations that represent different realizations of equivalent initial conditions in order to characterize the typical run-to-run variation of mass segregation and structural concentration. 

\begin{table*}
\centering
\caption{Summary of N-body simulations. For each simulation (identified by a unique ID) we report (from left to right) the initial number of stars; the initial IMBH mass in $M_\odot$; the velocity dispersion of the natal kick imparted to stellar remnants $\sigma_k$ normalized to the initial cluster velocity dispersion $\sigma_*$; the fraction of primordial binaries $f$; the metallicity $Z$ and the number of distinct realizations of the same initial conditions ($N_\mathrm{sim}$). Finally, each simulation is assigned to a specific sub-GROUP, which indicates, through a self-explanatory label, the main parameter that was varied with respect to the canonical initial conditions.}
\label{tab:simul}
\begin{tabular}{rlcccccl}
\hline
\text{ID} & \text{GROUP} & \text{N} &$M_{\mathrm{bh},0}$ &$\sigma_k/\sigma_*$ &$f$ &$Z$ &$N_\mathrm{sim}$\\
\hline
\text{can50k}   &\text{can}		    & 50k	&-		& 1.0	&-		&0.002		&4\\
\text{fb1050k}  &\text{bin}		    & 50k	&-		& 1.0	&0.10	&0.002		&1\\
\text{IMBH50k}  &\text{imbh}		& 50k	&100	& 1.0	&-		&0.002		&1\\
\text{Z50k}	    &\text{low-met}		& 50k	&-		& 1.0	&-		&0.001		&2\\
\text{kick50k}  &\text{high-kick}	& 50k	&-		& 2.0	&-		&0.002		&1\\
\hline
\text{can100k}  &\text{can}		    & 100k	&-		& 1.0	&-		&0.002		&7\\
\text{fb01}     &\text{bin}			& 100k	&-		& 1.0	&0.01	&0.002		&1\\
\text{fb03}	    &\text{bin}	    	& 100k	&-		& 1.0	&0.03	&0.002		&1\\
\text{fb05}     &\text{bin}			& 100k	&-		& 1.0	&0.05	&0.002		&2\\
\text{fb07}     &\text{bin}			& 100k	&-		& 1.0	&0.07	&0.002		&1\\
\text{fb10}	    &\text{bin}		    & 100k	&-		& 1.0	&0.10	&0.002		&1\\
\text{imbh}	    &\text{imbh}		& 100k	&100	& 1.0	&-		&0.002		&1\\
\text{IMBH}	    &\text{imbh}		& 100k	&200	& 1.0	&-		&0.002		&1\\
\text{Z}	    &\text{low-met}	    & 100k	&-		& 1.0	&-		&0.001		&1\\
\text{kick}	    &\text{high-kick}	& 100k	&-		& 2.0	&-		&0.002		&1\\
\text{kickfb05}  &\text{bin \& high-kick}   & 100k	&-		& 2.0	&0.05	&0.002		&1\\
\text{kickfb10}  &\text{bin \& high-kick}   & 100k	&-		& 2.0	&0.10	&0.002		&1\\
\hline
\text{can200k}	&\text{can}	        & 200k	&-		& 1.0	&-		&0.002		&1\\
\text{fb10200k}	&\text{bin}		    & 200k	&-		& 1.0	&0.10	&0.002		&1\\
\text{IMBH200k}	&\text{imbh}	    & 200k	&400	& 1.0	&-		&0.002		&1\\
\text{Z200k}	&\text{low-met}		& 200k	&-		& 1.0	&-		&0.001		&1\\
\text{kick200k}	&\text{high-kick}	& 200k	&-		& 2.0	&-		&0.002		&1\\
\hline
\end{tabular}
 \end{table*}
 
\subsection{Structural concentration index} \label{subsec:concentr_index}

In this work we introduce a novel definition for the concentration of a star cluster that can be readily applied to observations and is based on the integrated light profile. Our structural concentration index $R_5/R_{50}$ is defined as follows.

For each snapshot in our simulations, we consider a random projection in two dimensions and then restrict the analysis to the luminous component as identified by main-sequence stars only, including stars with mass $M \geq M^\mathrm{MS}_\mathrm{min}$, where generally $M^\mathrm{MS}_\mathrm{min}=0$ (i.e. no lower cutoff is applied). 

Considering only the main sequence stars that pass the mass cutoff, the center of the system is determined through a two step iteration: first, we compute the center of light of the system and then identify a first estimate of the projected half-light radius $R_\mathrm{50}$ (i.e. the radius that encloses half of the total light); second, we restrict the cluster to the region inside $2R_\mathrm{50}$ and take the new center of light as the final center of the cluster.

Finally, we calculate the concentration index as the ratio $R_5/R_\mathrm{50}$ which is given by the projected radii enclosing 5\% and 50\% of the light respectively. This is our standard definition for the structural concentration, which has the following advantages:
\begin{itemize}
    \item It is robust against confusion of low-mass stars in actual observations, as it only requires to mask effectively the light from stars brighter than main-sequence turn-off. This makes it potentially broadly applicable for galactic and nearby extragalactic globular clusters;
    \item Unlike the classical definition of concentration, which is the ratio of the core to the tidal radius of the system, it does not require fitting the surface brightness profile with a model to infer core and tidal radius. Therefore, it is easier to implement in the analysis of observational data, and does not suffer from possible systematic biases induced by the specific algorithm used for the King model fit.
 \end{itemize}

Finally, we also adopt a mass-based approach with the aim of testing the robustness of our results as well as mimicking high-quality observations in which stars are resolved individually. In this case, in addition to the standard analysis with $M^\mathrm{MS}_\mathrm{min}= 0$, we also apply a non-zero lower cutoff for main sequence stars, which is chosen as representative of state-of-the-art Hubble Space Telescope observations (see e.g. \citealt{libralato:18}). Also, the center of the system is determined using the center of mass and the structural concentration $R_5/R_{50}$ is calculated using radii that enclose a fraction of the total projected mass instead of light.

\subsection{Mass segregation} \label{subsec:mass_segr}

Different definitions have been proposed to quantify the degree of mass segregation of a stellar system. They can be broadly divided into two main approaches, with focus either on measuring the difference in mass at two given radii typically considering only main sequence stars \citep{gill:08}, or on the difference in the radial distributions of low versus high mass objects \citep{alessandrini:16,weatheford:17}. For the latter, generally the "low mass" population is identified with main sequence stars, while the high-mass objects are giants or blue stragglers.

Given the potential introduction of Poisson noise in the measurement from the lower number of giants/blue stragglers in both simulated and observed GCs, we adopt the approach of measuring the difference in average main sequence mass at different cluster radii. Specifically, we define the mass segregation indicator $\Delta m$ as: 
\begin{equation}\label{eq:delta_m}
  \Delta  m = \langle m \rangle_0 - \langle m \rangle_\mathrm{h},
\end{equation}
where $\langle m \rangle_0$ is average main-sequence stellar mass calculated at the center of the system (i.e. inside $0.1R_\mathrm{50}$) while $\langle m \rangle_\mathrm{h}$ is calculated for main sequence stars in the radial interval $[0.8R_\mathrm{50}:1.2R_\mathrm{50}]$. While our canonical approach is to consider light-based radii, when the concentration is computed with a mass-based approach, a consistent definition for $R_{50}$ as the projected half-mass radius is also used. Also, the low-mass cut-off $M^\mathrm{MS}_\mathrm{min}$ used for computing the concentration is self-consistently adopted for the mass segregation analysis. Finally, we note that we introduce a slightly idealized treatment of the simulations, motivated by computational convenience, and we assume we can resolve single masses in binary systems.

\subsection{Dynamical evolution overview} \label{subsec:dyn_evol}

Due to the cumulative effects of two-body encounters, star clusters experience dynamical relaxation on a time-scale comparable with the half-mass relaxation time $t_\mathrm{rh} = 0.138 N r_\mathrm{hm}^{3/2}/\log(0.11N)$ \citep{spitzer}. 
In addition, stellar evolution also has an impact on dynamics, as stars lose mass due to winds and explosive end-of-life events that ultimately lead to production of compact remnants. The gas lost by stars is generally not retained in the shallow potential well of GCs, thus perturbing the virial equilibrium and promoting expansion of the system. This naturally introduces an additional time scale in the system, independent of the relaxation time. Because of the interplay between stellar evolution and stellar dynamics, the study of the cluster's dynamical state becomes more challenging to link to fundamental physical processes and basic order-of-magnitude modeling, yet the approach clearly delivers a more realistic modeling of actual GCs compared to earlier studies of mass segregation that included gravity only (e.g., \citealt{gill:08}, \citealt{pasquato:09}).

\begin{figure*}
\centering
\includegraphics[width=.8\textwidth]{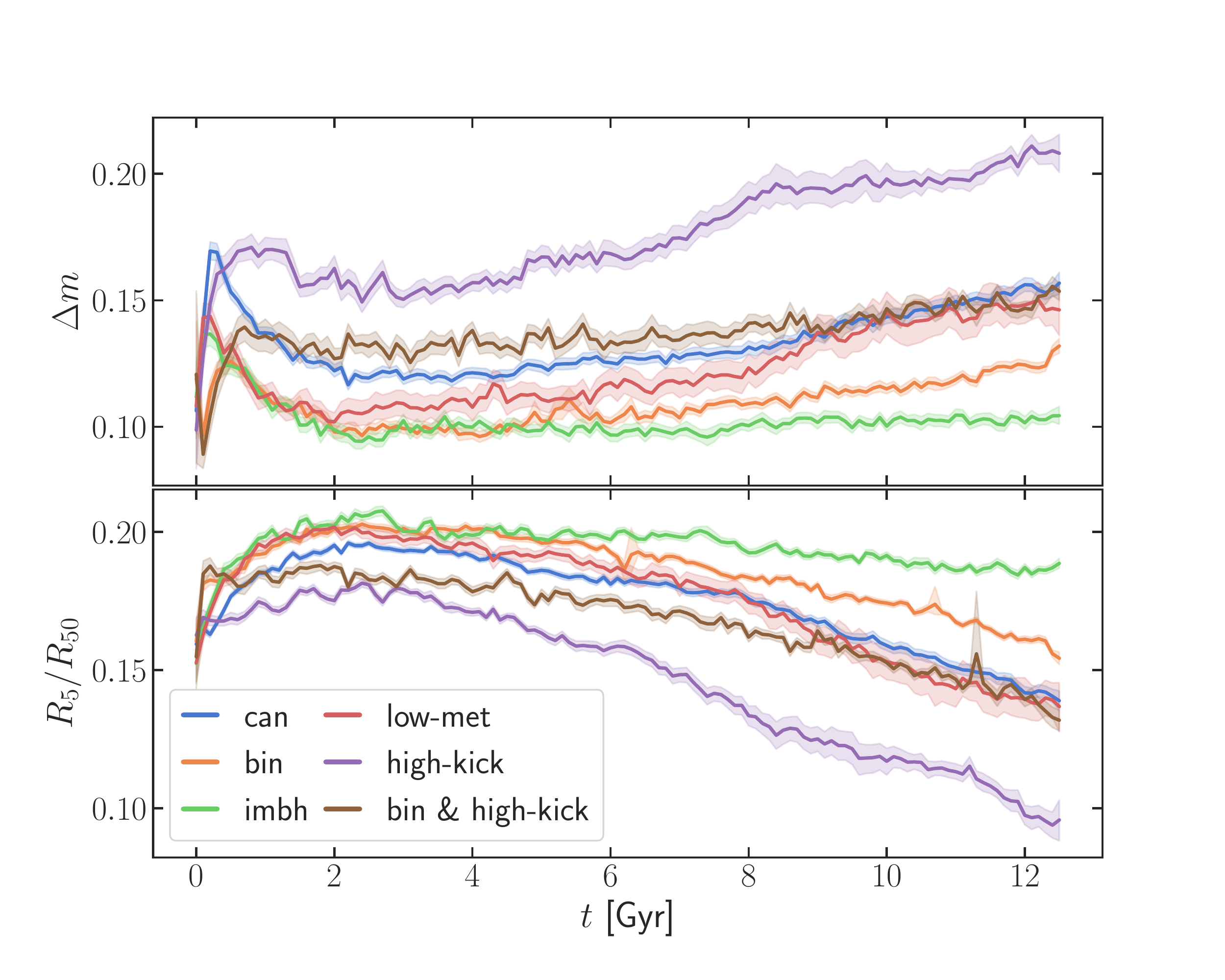}
\caption{Time evolution of mass segregation (top panel) and concentration (bottom panel) for different groups of simulations (see Table \ref{tab:simul}). At late times, the anti-correlation between the two quantities is apparent.}
\label{fig:ms_core_time}
\end{figure*}

In Fig.\ref{fig:ms_core_time}, we plot the entire evolution of mass segregation and concentration up until $12.5$ Gyr for selected groups of simulations. At early times, we can observe a rise in $\Delta m$ as the massive stars preferentially segregate toward the center, while the concentration decreases because of the expansion induced by mass loss due to stellar evolution. After a few $10^8$ yr, the main sequence turn-off has evolved to significantly lower masses, and thus the mass segregation indicator decreases in value. In turn, the stellar mass-loss rate decreases (the turn-off mass evolves slowly at later times), hence the cluster expansion decelerates and eventually contraction starts once two-body relaxation (gravothermal collapse) becomes the dominant evolutionary driver of the system. The system settles in a quasi-equilibrium long-term
evolution after the first $\sim 2$ Gyr, and a clear trend of correlation between concentration and mass segregation emerges
for all the different runs shown in the figure. The slow time evolution of $\Delta m$ in Fig.~\ref{fig:ms_core_time} is likely associated to steady mass loss from stellar evolution at late times, and was not observed in earlier studies, which reported instead a rapid settling of an equilibrium $\Delta m$ value after a few relaxation times in gravity-only simulations (\citealt{gill:08}, \citealt{pasquato:09}), thus highlighting the importance of including stellar evolution in the modeling. 

\section{Results: mass segregation - concentration correlation} \label{sec:results} 

The main result of our work is summarized in Fig.\ref{fig:ms_vs_core}, which investigates the relation between structural concentration and mass segregation for old star clusters. This plot highlights for the first time that a tight correlation between the level of mass segregation and the structural concentration index (as defined in subsection \ref{subsec:concentr_index}) exists for simulations characterized by a variety of different initial conditions. The figure has been obtained from all snapshots between 7.5 and 12.5 Gyr of age and includes all the simulations in Table \ref{tab:simul}, clearly showing that the more concentrated the cluster, the stronger the mass segregation. 

\subsection{Linear model}

To describe the correlation with the simplest meaningful model, we employ a linear model defined by
\begin{equation}\label{eq:master}
    \frac{\Delta m}{M_\odot} = a + b \frac{R_5}{R_\mathrm{50}}, 
\end{equation}
with $a$ and $b$ as free parameters. 

Their values ($a = 0.3246\pm0.0008$ and $b = -1.166\pm0.005$; see top entry in Table ~\ref{tab:fit}) are determined through a Markov-Chain Monte Carlo algorithm \citep{pymc3} as follows. We consider the targeted values of mass segregation as normally-distributed with standard deviation $\sigma=0.1$ and an expected value that is a linear function of the concentration index through Equation \eqref{eq:master}. Then, we evaluate the posterior distributions for the free parameters of the model, and determine the best-fit values together with the associated uncertainties from the mean, 2.5-th and 97.5-th percentiles of such distributions (see Fig.\ref{fig:best_fit}).

In addition to the linear model, we also performed a fit using a power-law, which is defined as in Equation \eqref{eq:master}, but with an extra free parameter as exponent of the concentration index. The best-fitting power law returned is close to linear, and the Akaike information criterion\footnote{According to the definition $\mathrm{AIC} = −2 \ln\mathcal{L}_\mathrm{max} + 2k$,
where $\mathcal{L}_\mathrm{max}$ is the maximum likelihood from the fit of a model with $k$ degrees
of freedom, the best model is the one that minimises AIC.} (see e.g. \citealt{liddle:07}) applied to the two different best-fit models indicates that the linear relation (AIC = 24.39) is slightly favored to describe our data compared to the power-law relation (AIC = 26.34). Hence, we consider the linear model as the preferred choice to describe the relation between mass segregation and structural concentration. 

\subsection{Time dependence}

In order to measure the light-based quantities in Equation \eqref{eq:master}, we rely on the complete set of simulations in Table \ref{tab:simul}. For the specific case of Fig.~\ref{fig:ms_vs_core}, the data-points are obtained in advanced stages of the evolution, namely considering the time interval $\Delta t$ between 7.5 and 12.5 Gyr. However, we also repeated the analysis at earlier times and for different sizes of the time interval (see Table \ref{tab:fit}). We find that a change of $\Delta t$ has a marginal effect on the best-fit parameters and root mean square error, with the largest relative difference below 20\%. This suggests that both mass segregation and concentration evolve with time in a strongly correlated fashion, with each simulation moving towards the upper left corner of Fig.~\ref{fig:ms_vs_core} remaining constrained along the linear relation \eqref{eq:master} (see also Fig.\ref{fig:ms_core_time}). 

Finally, we note that the spread in the level of mass segregation and structural concentration increases with time. This is evident from Fig.~\ref{fig:ms_core_time}, where the concentration index lies in the range 0.17-0.20 at 2 Gyr and in the range 0.10-0.19 at 12 Gyr (the same trend can be noticed for the degree of mass segregation). As a consequence, even though the correlation \eqref{eq:master} is not severely affected by the age of the cluster, the data-points in Fig.~\ref{fig:ms_vs_core} present a narrower distribution in both axes at earlier times, so that the differences in the degree of mass segregation and concentration among the various simulation groups are reduced.

\subsection{Correlation robustness}\label{sec:robustness}

To further investigate the robustness of our results, we repeated the analysis adopting mass-based definitions of the concentration. As we restrict our study to MS stars only, we find that using mass-based quantities has no significant impact on the best-fit parameters (relative differences within 15\%) but it increases the quality of the fit (lower RMSE). We interpret this as consequence of lower impact of shot noise in the definition of $R_5/R_{50}$, because light-based analysis is effectively carried out using only a small number of tracers (the most massive among the main sequence stars) due to the highly non-linear relation between stellar luminosity and mass ($L \approx M^4$). 

In contrast, we notice that the choice of a lower cut-off in the MS significantly affects the parameters of the linear relation \eqref{eq:master} (see Table \ref{tab:fit}).
The direct effect of changing $M^\mathrm{MS}_\mathrm{min}$ is to decrease the dynamic range for main sequence mass measurements. This leads to an increase of the average stellar mass, and to a decrease of the value of mass segregation measured. Instead, the structural concentration index is not sensitive to an increase of $M^\mathrm{MS}_\mathrm{min}$ (at least to first approximation), hence if the measure of $\Delta m$ decreases, then the linear relation between concentration and mass segregation becomes flatter.

\subsection{Structural concentration index versus classical King model definition}
Several definitions have been proposed in the literature to quantify the concentration in GCs (see e.g. \citealt{goldsbury:13}). One of the most widely used is the ratio of the truncation radius to the core radius as defined by \cite{king:62}. The core and truncation radii are obtained by fitting the surface density (or luminosity) profile with an empirical law (see Equation 14 in that paper)\footnote{Note that in case of large truncation radii, this definition of the core radius is equivalent to the radius at which the surface density (or luminosity) profile is equal to half of its central value.}. When we repeat our analysis using the core radius instead of $R_5$, we find that the root-mean square error of the best-fit linear model significantly increases, primarily due to a clear systematic deviation from a linear relation at low values of the concentration. This further motivates and justifies the approach of resorting to the structural concentration index defined in Subsection \ref{subsec:concentr_index} for future observational applications of the correlation.

\subsection{Impact of different dynamical constituents on mass segregation}

\begin{figure*}
\centering
\includegraphics[width=1.\textwidth]{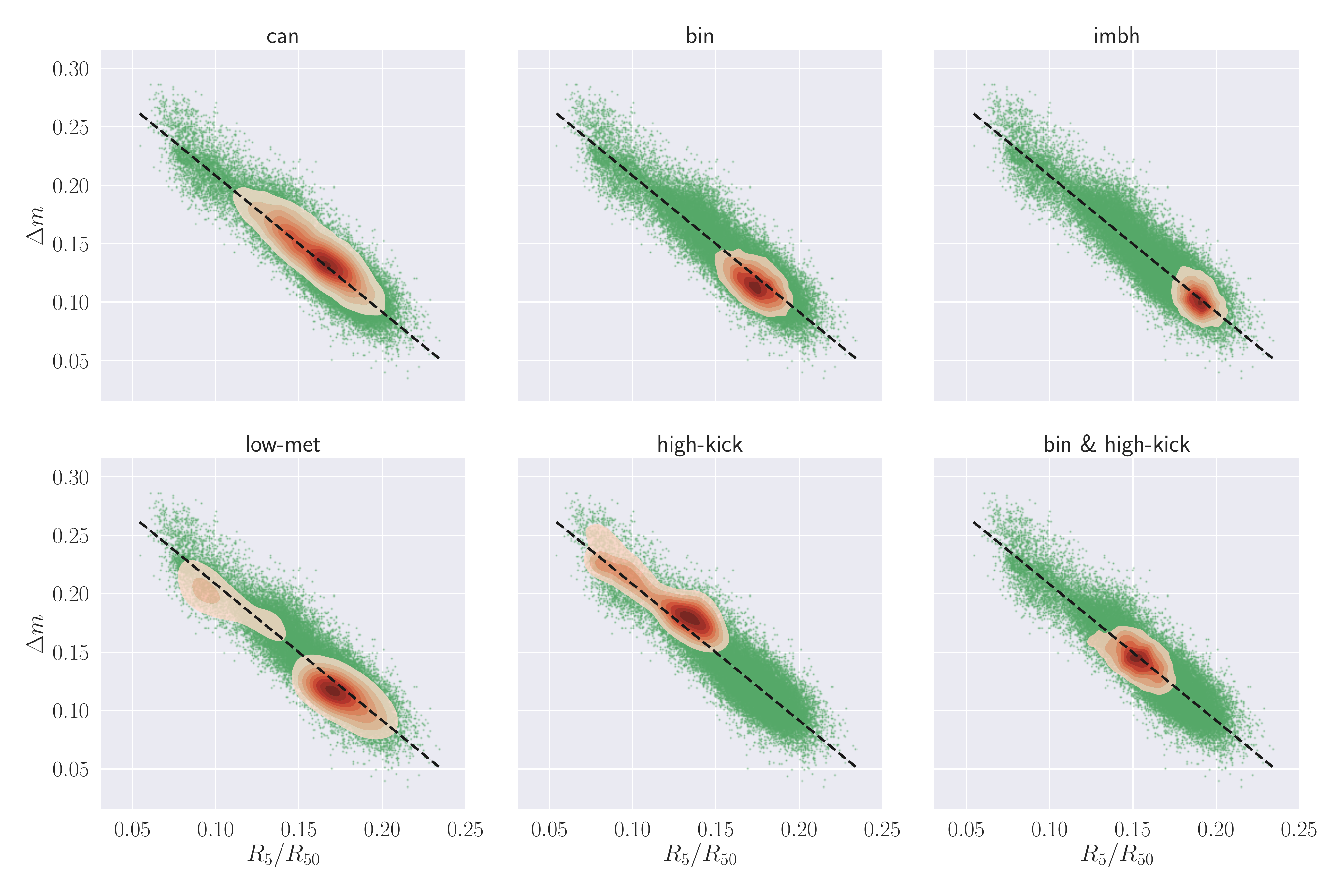}
\caption{Correlation between the degree of mass segregation $\Delta m$ and the concentration index $R_5/R_\mathrm{50}$ for clusters older than 7.5 Gyr. In each panel the probability density function of a single simulation group (see Table \ref{tab:simul} for definitions) is over-plotted as red shaded contours against the data points from all the available simulations (green dots) and the best linear fit expressed by Eq. \eqref{eq:master} (black dashed line).}
\label{fig:ms_vs_core}
\end{figure*}

\begin{figure*}
   \centering
   \includegraphics[width=.49\textwidth]{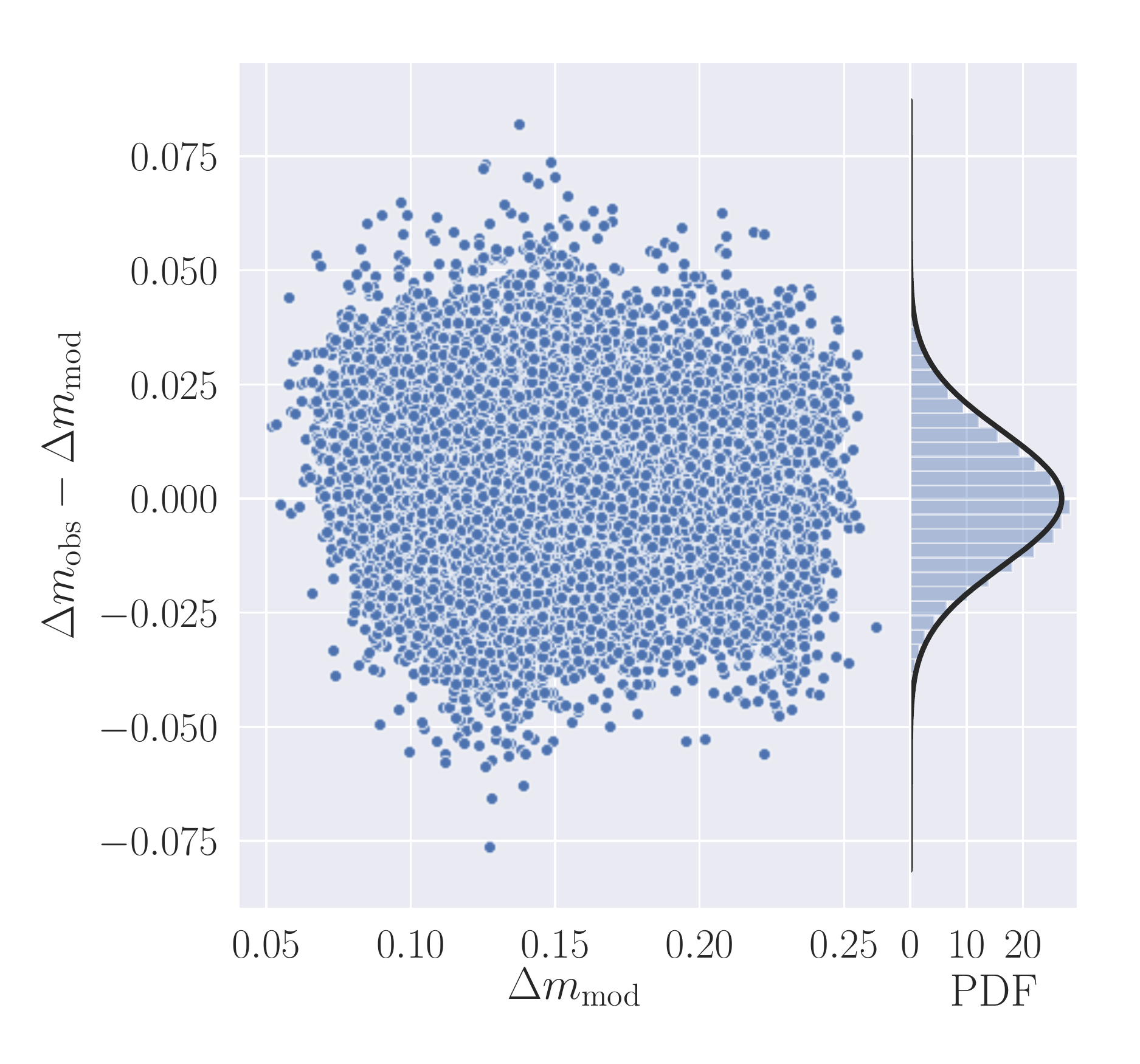}%
   \hspace{.01\textwidth}
   \includegraphics[width=.465\textwidth]{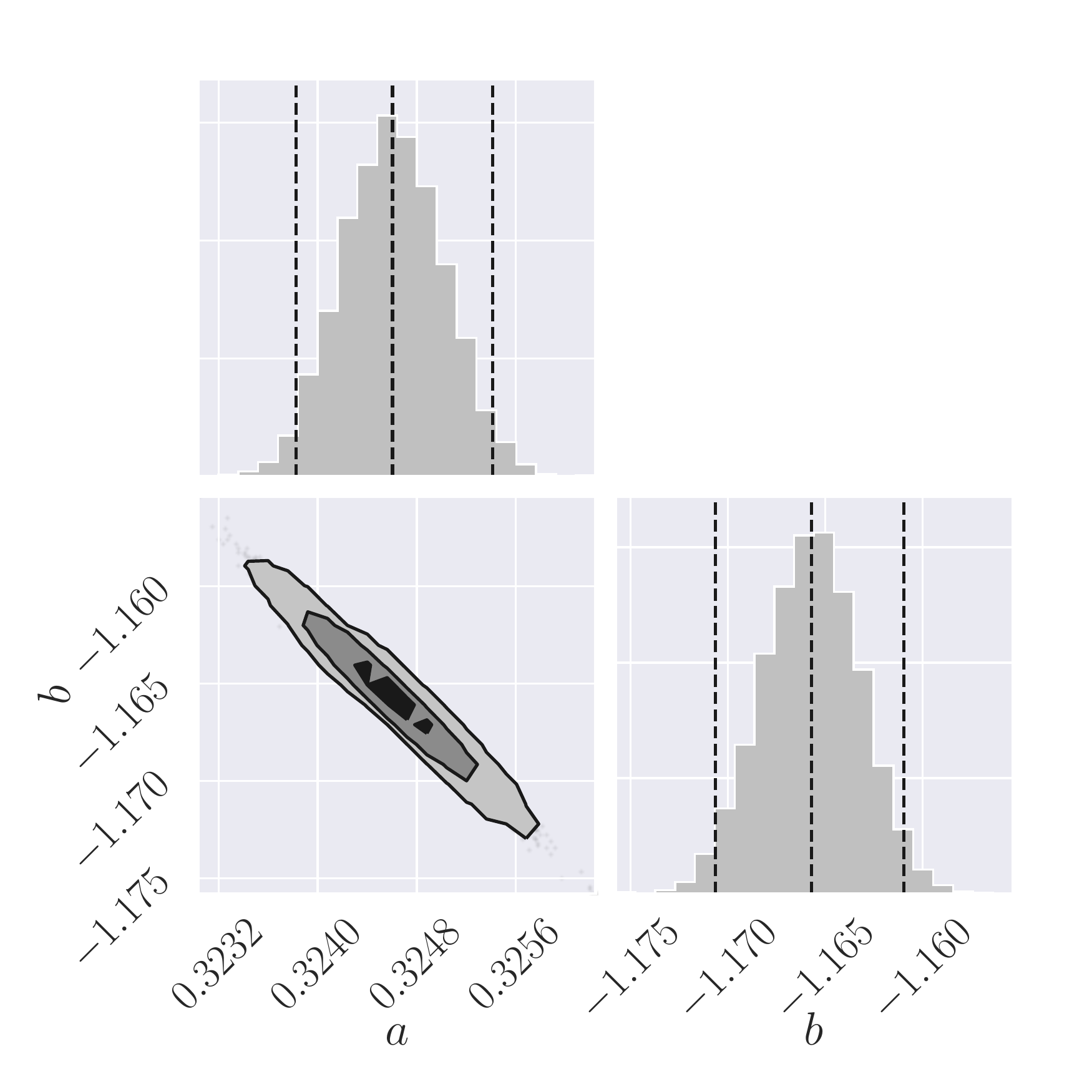}%
   \caption{Best-fit performance. In the left panel, we plot the residuals, i.e. the difference of the mass segregation's values measured in our simulations and predicted by model \eqref{eq:master}, against the predicted values. We also plot the residuals' probability density distribution (PDF) and its best-fit gaussian distribution (black line). In the right panel, we show the posterior distributions for the fitting parameters of the linear model as obtained from a MCMC analysis of our simulation sample.} 
   \label{fig:best_fit}
\end{figure*}

The panels in Fig.\ref{fig:ms_vs_core} clearly show that different groups of initial conditions are populating different regions of the structural-concentration versus mass segregation correlation. The general trends previously noted in the literature are recovered in our study. In particular, simulations with an IMBH are characterized by both a low concentration and a low amount of mass segregation when compared to the canonical case. A similar trend is present in some of the simulations with massive stellar BHs (low metallicity ICs), while simulations where dark remnants are given large natal kicks (hence a smaller likelihood of being retained) have a higher level of mass segregation. Simulations with primordial binaries show suppression of mass segregation as well, and appear to have a slight offset from the best fit. 

Further quantitative characterization of how the dynamical constituents of the system affect mass segregation and structural concentration is left to a future study.

\begin{table*}
\centering
\caption{Best-fit parameters corresponding to different observational and dynamical prescriptions used to calculate the concentration index and the amount of mass segregation for the model in Equation \eqref{eq:master}. For each linear model we indicate the approach used for the analysis (either light-based or mass-based); the lower mass cut-off of main sequence stars $M_\mathrm{min}^\mathrm{MS}$ in $M_\odot$; the time interval over which the linear relation is considered $\Delta t$ in Gyr; the intercept ($a$), slope ($b$), and root mean square error (RMSE) of the linear fit. The first row in this table represents our canonical light-based analysis.} 
\label{tab:fit}
\begin{tabular}{rrrrrr}
light &$M_\mathrm{min}^\mathrm{MS}$  & $\Delta t$ &$a$ &$b$ &RMSE \\
\hline
            yes            &-      &7.5-12.5   &$0.3246\pm0.0008$  &$-1.166\pm0.005$     &0.0148 \\
\hline
            yes             &-      &5.5-10.5   &$0.3437\pm0.0011$  &$-1.239\pm0.006$     &0.0161 \\
            yes             &-      &10-12.5    &$0.3201\pm0.0009$  &$-1.165\pm0.006$     &0.0133 \\
             no             &-      &7.5-12.5   &$0.3407\pm0.0007$  &$-1.285\pm0.004$     &0.0096 \\
             no             &-      &5.5-10.5   &$0.3444\pm0.0008$  &$-1.331\pm0.004$     &0.0095 \\
             no             &-     &10.0-12.5   &$0.3337\pm0.0009$  &$-1.227\pm0.006$     &0.0096 \\
             no           &0.2      &7.5-12.5   &$0.2500\pm0.0006$  &$-0.935\pm0.003$     &0.0085 \\
             no           &0.3      &7.5-12.5   &$0.1748\pm0.0005$  &$-0.649\pm0.003$     &0.0078 \\
\hline
\end{tabular}
 \end{table*}

 \section{Physical interpretation for the correlation} \label{sec:physic_interp}

To understand the physical origin of the mass segregation-structural concentration correlation expressed by Equation \eqref{eq:master}, we develop a simplified dynamical model, which we also use to  qualitatively explain the values of the best-fitting parameters derived from the data. For simplicity, we consider mass-based quantities. 

First, we assume that the cluster particle distribution at late times can be described by a self-similar isothermal distribution. Under this assumption, the surface density $\Sigma$ scales with the inverse of the projected radius (see Equation 4.105 in \citealt{binney}), so that the number of stars within a radial interval [0, $R$] is
\begin{equation}\label{eq:order_mag_1}
    N(R) \propto \int_0^{R} \Sigma(R') R' dR' \propto R.
\end{equation}
The total mass inside the same region is given by
\begin{equation}\label{eq:order_mag_2}
    M(R) \propto  \langle m \rangle_R N(R) \propto \langle m \rangle_R R,
\end{equation}
where $\langle m \rangle_R$ indicates the average stellar mass inside the projected radius $R$. Hence, we can derive that the 5\% lagrangian radius $R_5$ can be defined as: 
\begin{equation}\label{eq:R5model}
R_5 \propto M(R_5)/  \langle m \rangle_0,
\end{equation}
where $M(R_5)$ is 5\% of the total main-sequence mass of the cluster and we have approximated the average mass inside $R_5$ with $\langle m \rangle_0$ (the central value as defined in Sec~\ref{subsec:mass_segr}). Similarly, we can define the relation between half-mass radius and cluster half mass, and from that we can write: 
\begin{equation}\label{eq:order_mag_3}
   \frac{R_5}{R_{50}} \propto  \frac{\langle m \rangle_h}{\langle m \rangle_0}.
\end{equation}
Note that for an isothermal sphere without mass segregation, we would expect $R_5/R_{50}=0.1$, while due to the presence of a flat core in actual clusters the value of $R_5$, which is generally close to the observational core radius, is increased to $\sim 0.3 R_{50}$ (see, e.g., the observed core-to-half-light radius ratio in Fig.~3 of \citealt{trenti:10}, which is a close proxy).
After such empirical calibration, we can use the equation above to understand the impact of mass segregation through Eq.~\eqref{eq:delta_m} and a linear expansion:
\begin{equation}\label{eq:R5R50_OOMmodel}
   \frac{R_5}{R_{50}} \propto \frac{\langle m \rangle_h}{\langle m \rangle_h + \Delta m} \approx 0.3 \left( 1 - \frac{\Delta m}{\langle m \rangle_h} \right),
\end{equation}
where the latter is a Taylor expansion assuming small values of the ratio $\Delta m/\langle m \rangle_h$.
Finally, considering that $\langle m \rangle_h \approx 0.4 M_{\odot}$ and rearranging the terms to follow the structure of Eq.~\eqref{eq:delta_m} gives:
\begin{equation}\label{eq:delta_m_model}
\frac{\Delta m}{M_{\odot}} \approx 0.4-1.33\frac{R_5}{R_{50}}.
\end{equation}
Despite the simplicity of the model, this equation explains both the sign (anti-correlation), i.e. negative $b$, and the order of magnitude of the best fitting parameters in Table~\ref{tab:fit} (mass-based and $M_\mathrm{min}^\mathrm{MS}=0$).

This model is not suited for detailed quantitative analysis, because it relies on an isothermal sphere density distribution and neglects changes to the gravitational potential and particle density distribution that are induced by mass segregation. However, it is still very useful as a guide for interpretation of the data inferred from the full N-body dynamics, reinforcing the confidence in the potential use of our structural concentration index as a proxy for mass segregation. Finally, the model also suggests through Eq~\eqref{eq:R5R50_OOMmodel} that it is a change in mass segregation that induces the slow evolution of the concentration at late times observed in Fig.~\ref{fig:ms_core_time}.

\section{Discussion and Conclusion} \label{sec:conclusions}

In the present work we highlight for the first time the existence of a tight correlation (see Equation \ref{eq:master}) between the degree of mass segregation and the concentration of dense stellar systems, as identified by proxies that can be readily compared to observations of Galactic GCs. 

We presented and analyzed a new set of direct N-body simulations that include a variety of initial conditions, with particular focus on varying those dynamical constituents that are expected to mainly affect the long-term evolution of mass segregation (i.e. primordial binaries, stellar BHs and putative IMBHs). We find that the set of simulations considered do not present any significant outlier with respect to the correlation found. 

In addition, we tested the robustness of the correlation against different observational prescriptions. The fit performance together with the best fit parameters for different setups is reported in Table \ref{tab:fit}. Among the different prescriptions, the low-mass cutoff due to the very crowded regions of GCs is the one with the most prominent impact on the parameters of the correlation. We showed how this can have a simple yet insightful explanation in terms of an order-of-magnitude model, which has been introduced in Section \ref{sec:physic_interp} to provide with a physical interpretation for the observed mass segregation-concentration correlation.

This work has important implications on the current understanding of the dynamical evolution of dense stellar systems like GCs. Because of the physical quantities adopted in our study, the linear relation \eqref{eq:master} can be tested and calibrated in real observations of Galactic GCs. 

Thus, this tool offers a valuable potential opportunity to infer the dynamical state of GCs through the measurement of a structural quantity like the concentration index. In fact, such quantity can be used as a proxy for mass segregation, which, in contrast, is signficantly more challenging to measure even under the optimal case of high-quality space-based imaging for close Galactic GCs. In future studies we plan to validate the correlation using such observations, as well as to further exploit our set of simulations to explore the use of mass segregation as a tool to infer the dynamical constituents of old stellar clusters. 

\section*{Acknowledgements}
We thank the anonymous referee for constructive and useful suggestions.
This work was partially supported by the A.A.H. Pierce Bequest at the University of Melbourne. M.M. is grateful for support for this work provided by NASA through Einstein Postdoctoral Fellowship grant number PF6-170169 awarded by the Chandra X-ray Center, which is operated by the Smithsonian Astrophysical Observatory for NASA under contract NAS8-03060.  This research made use of Astropy, a community-developed core Python package for Astronomy \citep{astropy}. Numerical simulations have been performed on HPC clusters at the University of Melbourne (Spartan) and at the Swinburne University (OzSTAR).





\bibliographystyle{mnras}
\bibliography{biblio} 


\end{document}